# Chiral magnetic excitations in FeGe films

Emrah Turgut[1], Albert Park[1], Kayla Nguyen[1], Austin Moehle[1], David A. Muller[1,2], and Gregory D. Fuchs[1,2]

1. School of Applied and Engineering Physics, Cornell University, Ithaca, NY 14853, USA
2. Kavli Institute at Cornell for Nanoscale Science, Ithaca, NY 14853, USA

**Abstract:**

Although chiral magnetic materials have emerged as a potential ingredient in future spintronic memory devices, there are few comprehensive studies of magnetic properties in scalably-grown thin films. We present growth, systematic physical and magnetic characterization, and microwave absorption spectroscopy of B20 FeGe thin films. We also perform micromagnetic simulations and analytical theory to understand the dynamical magnetic behavior of this material. We find magnetic resonance features in both the helical and field-polarized magnetic states that are well explained by micromagnetic simulations and analytical calculations. In particular, we show the resonant enhancement of spin waves along the FeGe film thickness that has a wave vector matching the helical vector. Using our analytic model, we also describe the resonance frequency of a helical magnetic state, which depends solely on its untwisting field. Our results pave the way for understanding and manipulating high frequency spin waves in thin-film chiral-magnet FeGe near room temperature.

1. Introduction

Understanding the static and dynamic magnetic properties of materials is a key to their incorporation in active spintronic devices. In the light of the recent proposals for power-efficient spintronic memory devices based in chiral magnetism [1,2], it is increasingly important to characterize chiral magnetic materials in scalably grown thin-film form. Although the resonant spin dynamics in chiral magnetic films are more complex than conventional ferromagnetic resonance in uniformly-magnetized ferromagnetic films, understanding and measuring chiral magnetic excitations enables physical insight into the magnetic states of these materials and it offers quantitative characterization of dynamical properties that are relevant to future magnetic technologies.

The noncollinear spin texture that appears in chiral magnets is a consequence of the Dzyaloshinskii-Moriya interaction (DMI), which presents at interfaces and in the volume of noncentrosymmetric materials with broken inversion symmetry [3–6]. One class in these materials is cubic B20 crystalline monosilicides and monogermanides of transition magnetic elements, e.g. MnSi, FeCoSi, and FeGe [7]. Although they are in the same symmetry group, these silicides and germanides have surprising and distinctive electronic and magnetic properties depending on pressure, temperature, electric and magnetic fields [8–11].

Among B20 compounds, FeGe has the highest critical temperature, 278 K, for ordered chiral spin textures [7,12]. FeGe also has -0.6% lattice mismatch with the Si [111] surface, enabling scalable thin film growth [13,14], particularly in comparison with the mismatch of -3% and -6% for MnSi and FeCoSi, respectively [15,16]. Furthermore, in recent computational studies [17,18], thin film and nanoscale confinement of FeGe has been shown to stabilize the creation of a magnetic skyrmion, a two-dimensional chiral spin texture with non-trivial topological order. These properties make also FeGe thin films attractive for emerging spintronic applications with chiral magnets.

Although B20 FeGe thin films have been the subject of intense theoretical and computational studies, experimental studies have been limited to reports of the topological Hall effect and polarized neutron scattering measurements [13,14,19–23]. Furthermore, recent Lorentz transmission electron microscopy and transport studies of FeGe and MnSi thin films have brought into question the common interpretation of the topological Hall effect as arising solely from a skyrmion lattice phase [13,19,24,25]. These studies point out that transport measurements of B20 films are hard to interpret unambiguously because electron skew scattering by complex helical spin structures also contributes a Hall effect signal [13,26]. These difficulties motivate the application of alternative characterization methods to help identify chiral magnetic states and quantify magnetic behavior in thin film materials.

Microwave absorption spectroscopy (MAS) is a powerful tool to probe magnetization in both conventional ferromagnetic and complex materials [27–30]. In MAS, resonant absorption of a microwave magnetic field depends on the magnetic properties and configuration. For example, ferromagnetic resonance has been used to characterize effective magnetization, the damping parameter, and even magnetic anisotropies [31,32]. Moreover, MAS has been used to show universality of helimagnon and skyrmion excitations in bulk B20s regardless of being a conductor or an insulator [33]. Microwave fields are useful not only for understanding chiral magnets, but they can also create a giant spin-motive force in chiral magnets [34,35].

Here we report an experimental, theoretical, and micromagnetic study of chiral magnetic excitations in FeGe thin films by waveguide MAS. First, we describe the growth of B20 crystalline thin FeGe films via magnetron sputtering, and systematic characterization of their physical and magnetic properties by X-ray diffraction, electron backscattering diffraction, and magnetometry. Then, using parameters extracted directly from magnetic characterization, we study spin wave and resonant excitations in the helical and field-polarized states using micromagnetic simulations. With this framework to understand resonance frequencies and spin-wave modes, we experimentally perform temperature and magnetic field dependent MAS. We find that although the field-polarized magnetic resonance can be described by Kittel's formula, the helical state magnetic dynamics are more complex. In particular, an important mode occurs when spin waves are excited along the film thickness with a wavelength that matches the helical period. In addition, we theoretically calculate the resonant frequencies of the helicoids and find that it depends strongly on the value of the critical magnetic field that unwraps helicoid into the field-polarized spin state.

This paper is organized as follows: In Sec. 2, we describe the growth and characterization of the FeGe film. We report magnetometry studies in Sec. 3, and present micromagnetic simulations in Sec. 4. Then, we discuss experimental measurements of microwave absorption spectroscopy in Sec. 5. We analytically calculate resonance dynamics in helical magnet in Sec. 6. Finally, we conclude in Sec. 7.

## 2. Film growth and characterization

FeGe thin films are co-sputtered from Fe and Ge targets onto the surface of undoped Si [111] wafers and annealed post-growth at 350 ˚C for 30 minutes to create the B20 crystalline phase. The films are then characterized by X-ray diffraction and electron back-scattering diffraction (EBSD). In Fig. 1a, we show an X-ray $\theta - 2\theta$ scan and large area detector measurements of a 176-nm-thick FeGe film, which was determined by cross sectional imaging with a scanning electron microscope [see SM]. The narrow and point-like diffraction peak at θ=33.1° indicates a high degree of alignment of FeGe unit cells with respect to the Si [111] substrate.

From the X-ray diffraction data in Fig. 1a, the lattice constants of FeGe and Si along the normal axis of the film are found 4.680(2) Å and 5.441(2) Å, respectively. The reported lattice constant of FeGe at 290 K is 4.701 Å [36]. Then, we calculate the volume of a FeGe unit cell adjacent to the Si substrate is 103.91 Å$^3$, slightly larger than the bulk value of 103.86 Å$^3$. To quantify the tension on the FeGe lattice for a thin film, we use the bulk modules of 130 GPa and its derivative of 4.7, and Murnaghan formula from Ref. [36,37]. We find the pressure is -62 MPa. The recent studies of single crystal bulk MnSi reported a change in the uniaxial anisotropy and significant modification of the skyrmion phase diagram by applying positive pressure on the scale of several tens of MPa, particularly in comparison with the robust helical phase [38,39]. This agrees with Barla et al.'s study of bulk FeGe in which a several GPa pressure is needed to modify the chiral spin ground state [40]. Although a complete understanding of the magnetic phase diagram of thin film FeGe under stress requires more comprehensive experimental and theoretical study, to the best of our knowledge, our X-ray diffraction analysis ensures to form the helical and field-polarized magnetic states in our films.

Next we characterize the FeGe grains using plane-view transmission electron microscopy and EBSD. The transmission electron micrograph shown in Fig. 1b was taken using a Tecnai-F20 at 200 kV electron energy. It shows that our films have both ordered and disordered grains. We further investigate the nanoscale crystal configuration of the grains with EBSD (Figs. 1c-1f). The top image in Fig. 1c shows the scanning electron micrograph of an 8 μm by 4 μm region of the sample. The second EBSD micrograph (Fig. 1d) shows a crystalline phase map of the same region, confirming that 99% of the grains have the B20 phase. EBSD mapping also reveals grains and holes due to the lattice mismatch. The -0.23 % lattice mismatch between unit cells at the growth temperature suggests a 204 nm of average grain size [see SM]. From Fig. 1d, we see wide range of grain sizes, but they are mainly between 200 nm and 400 nm, which is close to our estimate.

In addition, we show the crystalline orientation map with grains aligned to the Si [111] direction in yellow (Fig. 1e), which also agrees with the X-ray diffraction data. More than 95% of the FeGe film is aligned with respect to Si substrate. Our in-plane orientation analysis of grains (Fig. 1f) shows twinning of the FeGe grains, which are rotated either +30 or -30 degrees in the plane with respect Si [111] unit cell. These high-quality polycrystalline films with grain sizes larger than the helical lattice constant of FeGe (70 nm) [41] allow us to study chiral magnetism in B20 thin film materials.

3. **Magnetic properties**

In this section, we characterize the magnetic properties of our FeGe thin films. First, we measure the magnetic moment of our film as a function of an external magnetic field and temperature using a vibrating sample magnetometer (VSM). Our films have an easy-plane magnetic anisotropy evidenced by an out-of-plane magnetic saturation field that is four times larger than the one for in the plane, as shown in Fig. 2a. Additionally, we find that as the temperature decreases, the magnetic moment and the saturation magnetic field increases. We note that while the out-of-plane magnetic moment curves do not indicate an obvious magnetic phase change, the in-plane magnetization curves have a feature at 400 Oe that does not appear in the magnetic hysteresis of a conventional ferromagnetic material. To reveal the features better, in Fig. 2b, we plot the derivative of the in-plane magnetization with respect to the applied magnetic field. The emergence of a peak in the susceptibility below the critical temperature 273 K indicates a magnetic phase transition. Such magnetic phase changes are evidence of a transformation from an out-of-plane q-axis helical phase into a field-polarized phase by unwinding of the in-plane moment in accordance with the data in the following sections and previous polarized neutron scattering studies in FeGe and MnSi thin films [19,21].

### 4. Micromagnetic simulations

In this section, to identify the spin dynamics in FeGe thin films, we perform micromagnetic simulations using the Mumax3 software [42]. We identify the magnetic properties of the film as simulation parameter inputs using magnetometry measurements in Fig. 2 and the following relations: $H_k = \frac{\pi^2}{16} H_d$, where $H_k = 450\ Oe$ is the untwisting field and $H_d = 730\ Oe$ is the saturation field. The saturation field in chiral magnets is described by $H_d = \frac{D^2}{2AM_s} = \frac{8\pi^2 A}{L_D^2 M_s}$, where the saturation magnetization $M_s = 150$ kA/m and the helical period $L_D = 70\ nm$ [21,24,34,43]. We find $A = 6.8 \times 10^{-13}$ J/m. We also assume that the helical period $L_D = 70\ nm$ does not depend on the saturation magnetization or temperature [21]. In the simulation, the sample dimension is 3.2x3.2x176 nm³ in the *x, y,* and *z* directions, and the unit cell is a 0.8x0.8x0.8 nm³ cube. We apply 16 repetitive periodic boundary conditions in the *x* and *y* directions to mimic the uniform film [see SM for a sample input code]. Furthermore, we study chiral dynamics at 0 K, so we did not implement a fluctuating thermal field, which is necessary to quantitatively capture phase transitions between the helix and the field-polarized states. Thus, we supply this information to the simulation by initializing the magnetic states based on experimental magnetometry results.

In micromagnetic simulations, we use the ringdown method to obtain dynamic properties of spins. In the ringdown method, we first initialize the system in the helical state between -500 Oe and 500 Oe, and the field-polarized state for the rest of the magnetic fields. Then, for each field, we relax the system to its equilibrium state, where all torques vanish. For example, we show these equilibrium spin configurations at 1750 Oe, 250 Oe, and 0 Oe in-plane fields in Figs. 3b-d. Next, we apply a magnetic pulse with a Gaussian profile, and record the *x*, *y*, and *z* components of the local magnetic moments at 25 ps time steps for a 20 ns duration [see SM for details]. The Gilbert damping parameter is set to an artificially small value ($\alpha$=0.002) to capture enough periods of the natural oscillations so that we can identify the modes that are sustained by microwave driving [44,45]. To be consistent with the coordinate system we use for theoretical calculations in Sec. 6, we also perform a coordinate transformation of the magnetization components from Cartesian coordinates into spherical coordinates. The *z* component of magnetization simply becomes θ, whereas the azimuthal angle φ is calculated from the *x* and *y* components in the plane.

To calculate the natural modes and frequencies, we compute the discrete Fourier transform of the local magnetic deviation from equilibrium for each magnetic field. Because the deviation in both θ and φ angles results in the same resonance frequencies and modes, we plot only θ in Fig. 3. Next, we compute spatially-averaged Fourier coefficients of all spins to obtain the power spectral density (PSD) [45]. Fig. 3a shows the PSD for frequencies between 0.5 and 7 GHz, and in-plane magnetic fields between -2000 Oe and 2000 Oe. We identify three magnetic fields (H=1750 Oe, 250 Oe, and 0 Oe) to explore the resonance behavior and modes. We also define wrapping number ς, which is $(\varphi_N - \varphi_1)/2\pi$ total wrapping of spins at the equilibrium configuration. To reveal the modes, we plot the Fourier coefficients as a function of the frequency and thickness (*z* direction) in Figs. 3b-d, with the spin configurations along the thickness shown above each plot.

The first region is at H=1750 Oe, where we observe a Kittel-type uniform resonance of the field-polarized state at 5 GHz (Fig. 3d.) There are also edge modes which are inversely proportional to the magnetic field at 3 GHz. The second region is at H=250 Oe, where the spins wrap $\varsigma$ = 2.46 times. The resonance frequencies are located at 4.5 GHz, 2.6 GHz, and 0.5 GHz, and the corresponding number of nodes are 4, 2, and 0, with only even numbers because $2\varsigma$ = 4.9 < 5. On the other hand, in the third region at H=0 Oe, the system is driven into $\varsigma$ = 2.65 times wrapping, which is slightly larger than the expected 2.51 (176 nm /70 nm) because of the demagnetizing field of the film. The resonance frequencies then increase to 5.5 GHz, 3.0 GHz, and 0.8 GHz, and the number of nodes becomes odd–5, 3, and 1, respectively, because $2\varsigma$ = 5.3 >5. These results show sensitivity of spin waves in the spin configuration–helical wrapping in chiral magnet.

## 5. Experiment: Microwave absorption spectroscopy

After we account for spin waves in the helical and field-polarized states, we experimentally perform magnetic resonance measurements by placing FeGe film on a broadband metallic coplanar waveguide (CPW). More details about the design and characteristic of CPW can be found in Ref [46]. We apply RF field with a signal generator and monitor the transmitted power with a RF diode as a function of magnetic field and temperature. To remove any non-magnetic signals, we lock-in to the transmitted RF power referenced a magnetic field modulation that we introduce using an ac field coil. Thus, we measure the derivative of the transmitted power, $\Delta P/H_{ac}$, as shown in Fig. 4. For each temperature, we fix the sample temperature and vary the microwave frequency from 0.5 GHz to 7 GHz with a 0.25 GHz step size, and the magnetic field from 3000 Oe to -3000 Oe with a 30 Oe step size. The temperature of the sample is controlled by a Peltier element that allows a convenient temperature control between 300 K and 255 K.

In Fig. 4a and b, we plot the microwave absorption spectra at 285 K and 258 K, respectively (data for the full temperature range can be found in SM). At 285 K, the film is in a paramagnetic state, whereas it can be in either the helical or field-polarized state at 258 K. While, the paramagnetic state shows hardly any absorption, we find two important resonances in FeGe films from Fig. 4b. The first is a uniform, field-polarized magnetic resonance that is well-described by Kittel's formula. We extract the resonance fields, frequencies $f$, and the linewidths $\Delta H$ [see SM]. By fitting linewidth to $\Delta H = \Delta H_0 + \frac{4\pi\alpha f}{\gamma_e}$, where $\gamma_e$ is the electron's gyromagnetic ratio (2.8 MHz/Oe) [47], we find the damping constant $\alpha$ is 0.038 ± 0.005 at 258 K. This α is substantially lower than the recently reported value of 0.28 in thinner FeGe films in an out-of-plane magnetic field applied along the [111] orientation and at unreported temperature [45].

The second resonance is the helical resonance, also known as the helimagnon. As we point out by a vertical arrow in Fig. 4b, the helical resonance has a narrow field range (450-500 Oe) but a wide frequency range (4-5 GHz), in contrast to the field-polarized phase. We attribute this experimental observation to the helical resonances that we described in the second region of the micromagnetic simulation that appeared at 4.5 GHz (Fig. 3c). The wavelength of the 4.5 GHz spin wave matches to the helical period of the FeGe film, which shows how spin waves are explicitly filtered by the helical spin texture in B20 thin films. In other words, two constraints: the thickness and the helical vector, impose a specific discretization of the spin wave spectrum in our films.

We also plot MAS as a function of magnetic field and temperature at a constant frequency of 4.5 GHz (Fig. 4c.) It is interesting to note that the field-polarized resonance extends up to 280 K, whereas the helical resonance disappears at temperature above 265 K, which is close to the critical temperature of FeGe. This finding agrees with our magnetometry measurements that show some magnetic moment persists above the critical temperature 273 K. This precursor magnetic region between 273 and 280 K was also observed by Wilhelm *et al.* in the magnetic susceptibility measurements of the bulk FeGe [12]. However, Wilhelm *et al.* found a relatively small precursor region only between 278 K and 280 K. From the magnetic susceptibility measurements [see SM], we find a critical temperature 273 K for the helical order, which is lower than the bulk crystal value of 278.2 K. Furthermore, Wilhelm *et al.* observed a skyrmion phase only between 273 K and 278 K [12], which lies above the helical ordering temperature in our films. Therefore, it is an open question as to whether the skyrmion phase would shift to lower temperatures or totally disappear in thin films.

To better understand the helical resonance, we locate the peak of the resonance in field and frequency and track it as a function of temperature. In Fig. 4d, we observe a monotonic decrease of the resonance field and frequency as the temperature increases. This is consistent with our magnetometry measurements (Fig. 2) that show both the magnetic moment of the film and the unwrapping critical magnetic field decrease with increasing temperature. Such a decrease in the helical-phase resonance frequency and increase in the field-polarized phase resonance frequency for increasing temperature was also observed by Schwarze *et al.* in bulk B20 materials [33].

In contrast to observations in bulk crystal [33] and micromagnetic simulations (Fig. 3), our experiment does not show any clear microwave absorption at 0 field. This difference may arise from the grain formation in our films, which are not accounted for in micromagnetic simulations. At H=0 Oe, we think, oscillations in twinned grains are irregular and do not show strong absorption, whereas a nonzero field helps unify the collective motion of spins perpendicular to the field, enabling a strong helical state resonance absorption before untwisting into the field-polarized state.

6. **Theoretical calculations**

By comparing the micromagnetic simulations and experimental measurements of MAS, we identified the resonance frequencies and spin-wave modes in the helical and field-polarized states. In this section, we also analytically model excited chiral helimagnets to account for spin wave excitations.

We describe the one dimensional Hamiltonian density of a chiral helimagnet by

$$\mathcal{H} = A(\partial_z \boldsymbol{m})^2 - \boldsymbol{D}.\boldsymbol{m} \times \partial_z \boldsymbol{m} - \boldsymbol{B}.\boldsymbol{m} + K_u(\boldsymbol{m}.\hat{\boldsymbol{n}})^2 - \frac{1}{2}\boldsymbol{H_m}.\boldsymbol{m}, \qquad (1)$$

where $A$ is the exchange stiffness constant, $\boldsymbol{D}$ is the DM interaction constant, $\boldsymbol{B}$ is the external magnetic field, $K_u$ is the anisotropy constant and $\boldsymbol{H_m}$ is the demagnetizing field due to shape of the sample. We use the normalized magnetization $m =$ $[\sin \theta(z,t) \cos \phi(z,t), \sin \theta(z,t) \sin \phi(z,t), \cos \theta(z,t)]$ in the spherical coordinate as in the previous section [21,34,48]. The external magnetic field **B** includes the dc field $B_x$ and the ac microwave field $B_y$, and it is written as $\boldsymbol{B} = [B_x, B_y \sin \omega t, 0]$. Next, we write down the Lagrangian density $\mathcal{L}$,

$$\mathcal{L} = -\frac{\hbar M_s}{g_e \mu_B}(\cos \theta - 1)\partial_t \phi - \mathcal{H}, \tag{2}$$

where $M_s$ is the saturation magnetization, $\mu_B$ is the Bohr magnetron, and $g_e$ is the electron g-factor. The equations of motion are constructed by expressing Eq. 1 and 2 in terms of $\theta(z,t)$ and $\phi(z,t)$ coordinates [see SM].

For the equilibrium helical state at **B**=0, the solutions are simply $\theta = \frac{\pi}{2}$ and $\phi = Qz$, where $Q = \frac{D}{2A}$ is the helix wave number. Application of an external magnetic field creates a deviation from equilibrium by $\theta_1$ and $\phi_1$ as

$$\phi(z,t) = Qz + \phi_1(z) \sin \omega t, \tag{3}$$

$$\theta(z,t) = \frac{\pi}{2} + \theta_1(z) \cos \omega t. \tag{4}$$

The distorted helix has been described by cosine expansions of the angles as in $\phi_1(z)[\theta_1(z)] = A_1[B_1] + A_2[B_2] \cos Qz + A_3[B_3] \cos^2 Qz$, where $A_{1\text{-}3}$ and $B_{1\text{-}3}$ are coefficients for $\phi_1$ and $\theta_1$, respectively [34]. As the last step, we substitute Eq. 3 and 4 into the equations of motion using the small angle approximation [see SM]. We also use the same material parameters obtained in the micromagnetic simulation section. Finally, we solve the eigenvalue problem for the resonance frequencies and modes.

In Fig. 5a, we plot the real part of the three resonance frequencies $f_1$, $f_2$, and $f_3$ as functions of in-plane field $B_x$. If we define a critical field $H_c$=830 Oe, where the solutions to $f_1$ and $f_2$ become degenerate (Fig. 5a): $Im[f_1] = Im[f_2] = 0$ and $Im[f_3] \neq 0$, for $H < H_c$: whereas $Im[f_1] \neq Im[f_2] \neq 0$ and $Im[f_3] = 0$, for $H > H_c$ [see SM]. Therefore, the solutions to $f_1$ and $f_2$ above $H_c$ does not support natural oscillations. Additionally, the small angle approximation and series expansion are only valid at the low field (in the pink-filled region of Fig. 5a), because the phase change from the helical into the field-polarized state happens at $H_d$ =730 Oe (Fig. 2.)

Resonance modes in $\theta_1$ at $f_1$ and $f_2$ are shown in Fig. 5b and 5d, and the ones in $\phi_1$ at $f_1$ and $f_2$ are shown in Fig. 5d and 5e, respectively. At $f_1$, there are five nodes that match the helical period, whereas the number of nodes double for $f_2$. This doubling of nodes is because of expansion of $\theta_1$ and $\phi_1$ up to the second order ($cos^2kz$.) When we increase in-plane $B_x$ field, the oscillation amplitudes for $f_1$ decrease (Fig. 5b and 5d) and the amplitudes for $f_2$ increase (Fig. 5c and 5e.) This suggests that an in-plane field drives the symmetric helimagnon into the distorted helimagnon by decreasing the *cos kz* term and increasing the *cos$^2$kz* term. The static distortion of helimagnets by an in-plane field was previously observed using a polarized neutron scattering experiment, which is in agreement with our findings from oscillation amplitudes [19].

Our analytic calculation does not take into account the thickness of the film, therefore we do not observe additional modes along the film thickness as we did in micromagnetic simulation. However, we confirm the increase of the resonance frequency by application of a larger magnetic field, in agreement with micromagnetic simulation (Fig. 3a). On the other hand, Schwarze *et al.* observed the opposite in the bulk B20, i.e. the larger the field, the smaller the frequency. An important difference, however, is that in our films, the magnetic field untwists the helix, while in their bulk crystal the magnetic field introduces a conical angle to the helical phase. Therefore, an opposite dependence to the magnetic field is consistent with our theoretical understanding. Nevertheless, our calculation predicts two resonance frequencies at $B_x$=0 regardless of the geometry. At $B_x$=0, the resonance frequencies become $f_1 = \frac{g_e H_d \mu_b}{2\pi\hbar}$ and $f_2 = \frac{\sqrt{10} g_e H_d \mu_b}{2\pi\hbar}$. For example, Schwarze *et al.* found resonance frequencies in the range of 14–17 GHz, 3–4 GHz, and 1.5–2 GHz for bulk MnSi, FeCoSi, and $Cu_2OSeO_3$, respectively [33]. The critical fields were also reported as 0.5, 0.1, and 0.04 T. From $f_1 = \frac{g_e H_d \mu_b}{2\pi\hbar}$ formula, we estimate $f_1$= 14 GHz, 2.8 GHz, and 1.1 GHz for each materials, respectively, in close agreement with observations. Therefore, our theoretical approach provides a straightforward estimate of helimagnon frequencies.

This relation to the critical field coincides with the recently developed microscopic theory of spin waves in cubic magnets with DMI by Maleyev [49]. He found that the spin-wave stiffness $D_{sw} = \frac{g_e \mu_B H_d}{Q^2}$, which is 0.105 eV Å² for our FeGe thin film. This value is in precise agreement with the recent neutron scattering experiment on bulk FeGe at 250K [50]. One can define the spin-wave resonance frequency in a chiral magnet by $f = f_{heli} + \frac{D_{sw}}{h}\left(\frac{\pi n}{L}\right)^2$, where $h$ is Planck's constant, $L$ is the thickness, $n$ is the mode number, and $f_{heli}$ is natural helical frequency which is found 2.0 GHz in our FeGe films using $f_1$ (Fig. 5a.) Therefore, the frequency becomes $f = 2.0 + n^2 0.081$ GHz. For n=3 and n=5, we find spin wave frequencies at 2.7 GHz and 4.0 GHz. These frequencies are close to the ones we observed in the micromagnetic simulations in Fig. 3b. Small differences may be originated from variations in the material parameters, because they are highly sensitive to the temperature around the critical temperature $T_c$. Our analytic approach is a direct and simple method to identify the resonance frequencies in chiral thin film magnets. Full understanding of spin-waves in confined chiral magnets will require further theoretical and experimental study.

## 7. Conclusion

In conclusion, we present a comprehensive experimental and theoretical study of the microwave resonance dynamics in a chiral magnetic FeGe thin film. We grew FeGe films by magnetron sputtering and systematically characterized their physical and magnetic properties. Our films are polycrystalline but have high-quality B20 crystal phase, confirmed by the electron backscattering diffraction. Below the critical temperature, static magnetometry measurements show that the film has a helical to field-polarized magnetic phase transition at 450 Oe under an in-plane magnetic field. Our microwave absorption measurements also show resonance features for the helical and field-polarized states. By comparing our experimental measurements with micromagnetic simulations and analytic calculations, we demonstrate that the helical state has resonant microwave dynamics that are highly sensitive to the twisting spin texture. Our analytical calculations also show that the resonance frequencies can be described by the untwisting critical magnetic field, which is in agreement with micromagnetics and experimental observations. Our results pave the way toward understanding spin wave dynamics in chiral and topological spin textures, grown as thin films without any limitation to scalability, thus promising for an integration of chiral spintronics.

**Acknowledgement**

This work was supported by the DOE Office of Science (Grant # DE-SC0012245). We also acknowledge use of facilities of the Cornell Center for Materials Research (CCMR), an NSF MRSEC (DMR-1120296), and CCMR research support for K.N. and D.M., who collaborated on transmission electron microscopy measurements. We further acknowledge facility use at the Cornell Nanoscale Science and Technology Facility (Grant # ECCS-1542081), a node of the NSF-supported National Nanotechnology Coordinated Infrastructure. We thank Robert M. Hovden and Lena F. Kourkoutis for valuable discussions and on-going help with electron microscopy, and F. Guo for reviewing the manuscript.

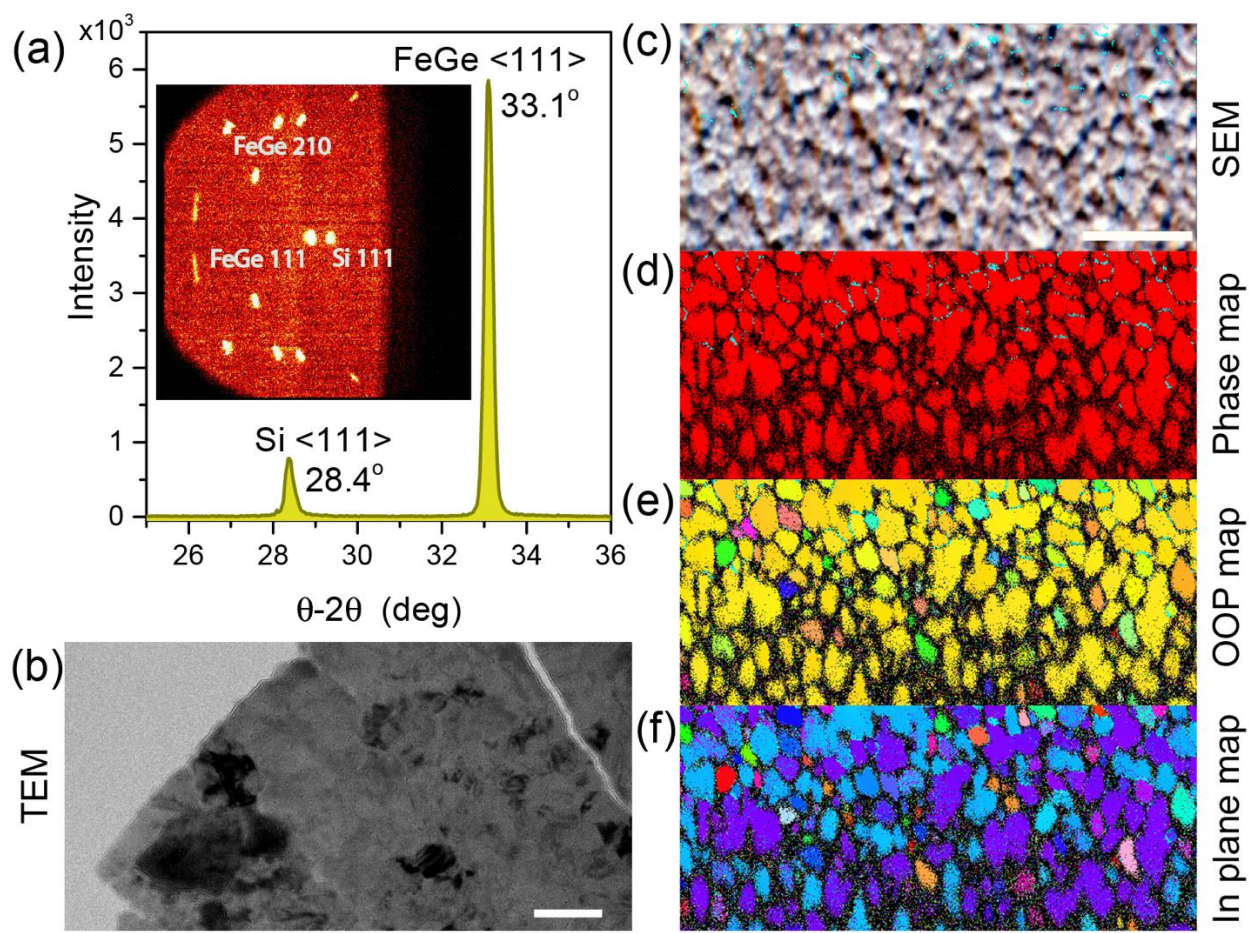

**FIG 1** X-ray and electron diffraction characterizations and transmission electron micrograph of FeGe thin film. a) shows θ-2θ scan of X-ray diffraction with an additional χ angle profile in the inset. Having sharp peaks instead of rings suggests good alignment of FeGe film. b) is transmission electron micrograph of plain-view of the film (scale bar is 100nm). c) is scanning electron micrograph, d) is the crystalline phase map of the same region indicating >99% B20 phase. e) is the out-of-plane alignment and f) is the in-plane alignment of the grain. The scale bar is 2 μm through (c)–(f).

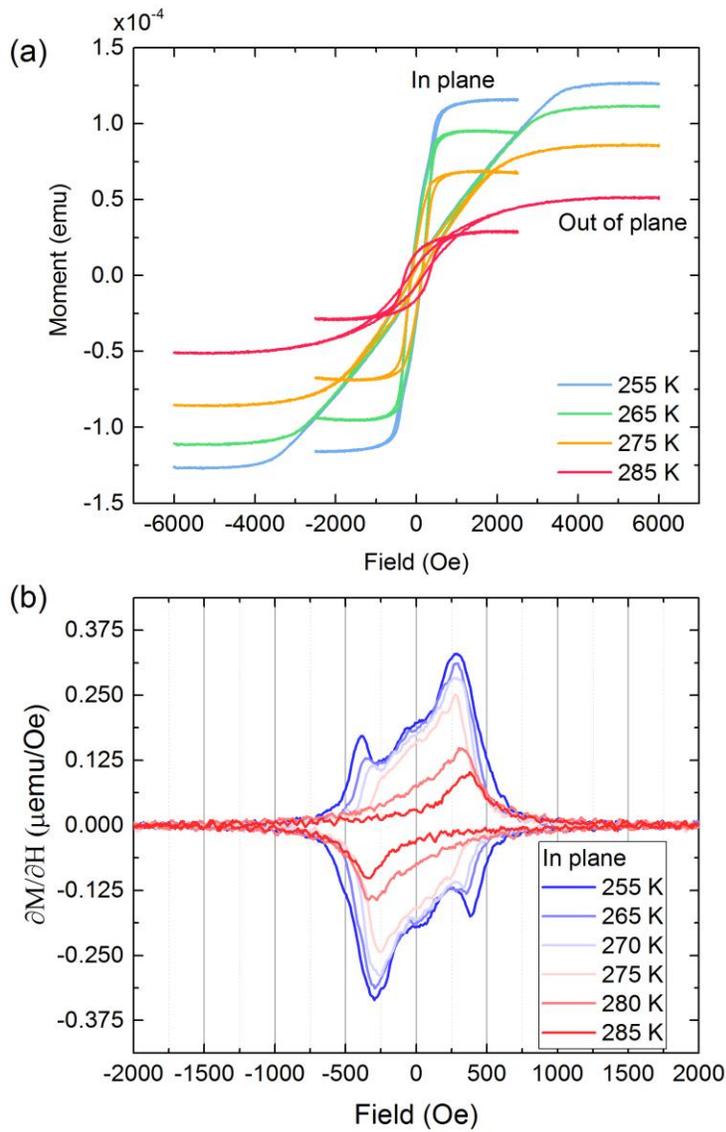

**FIG 2** Magnetometry measurements of our thin film FeGe. (a) shows M-H curves for both the in-plane and out-of-plane fields, and (b) shows the derivative of the in-plane magnetization with respect to the applied magnetic field to better reveal unwinding of the helical phase.

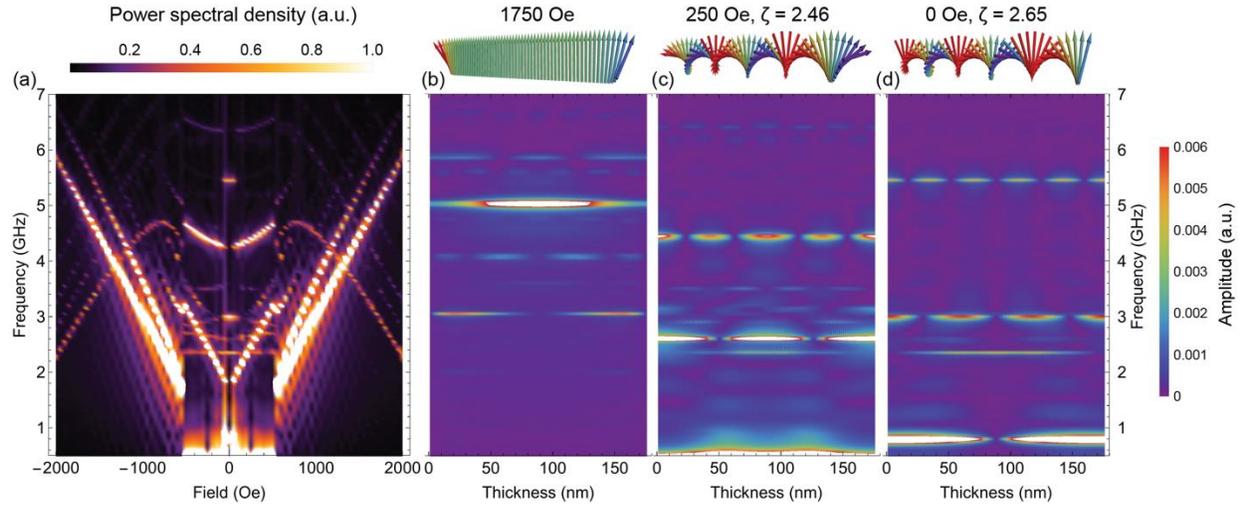

**FIG 3** Power spectral density (PSD) and natural oscillation modes of spin waves. a) shows the PSD of spatially summed Fourier coefficients as a function of in-plane magnetic field. The helical spin configuration presents between -500 Oe and 500 Oe, and the field-polarized state presents for the rest. b) at $B_x$ = 1750 Oe field, the field-polarized state has a Kittel-type uniform mode. c) at $B_x$ = 250 Oe field, the wrapping number $\zeta$ is 2.46, which allows only even nodes (0, 2, and 4.) d) at $B_x$ = 0 Oe field, the wrapping number $\zeta$ is 2.65, which allows only odd nodes (1, 3, and 5.)

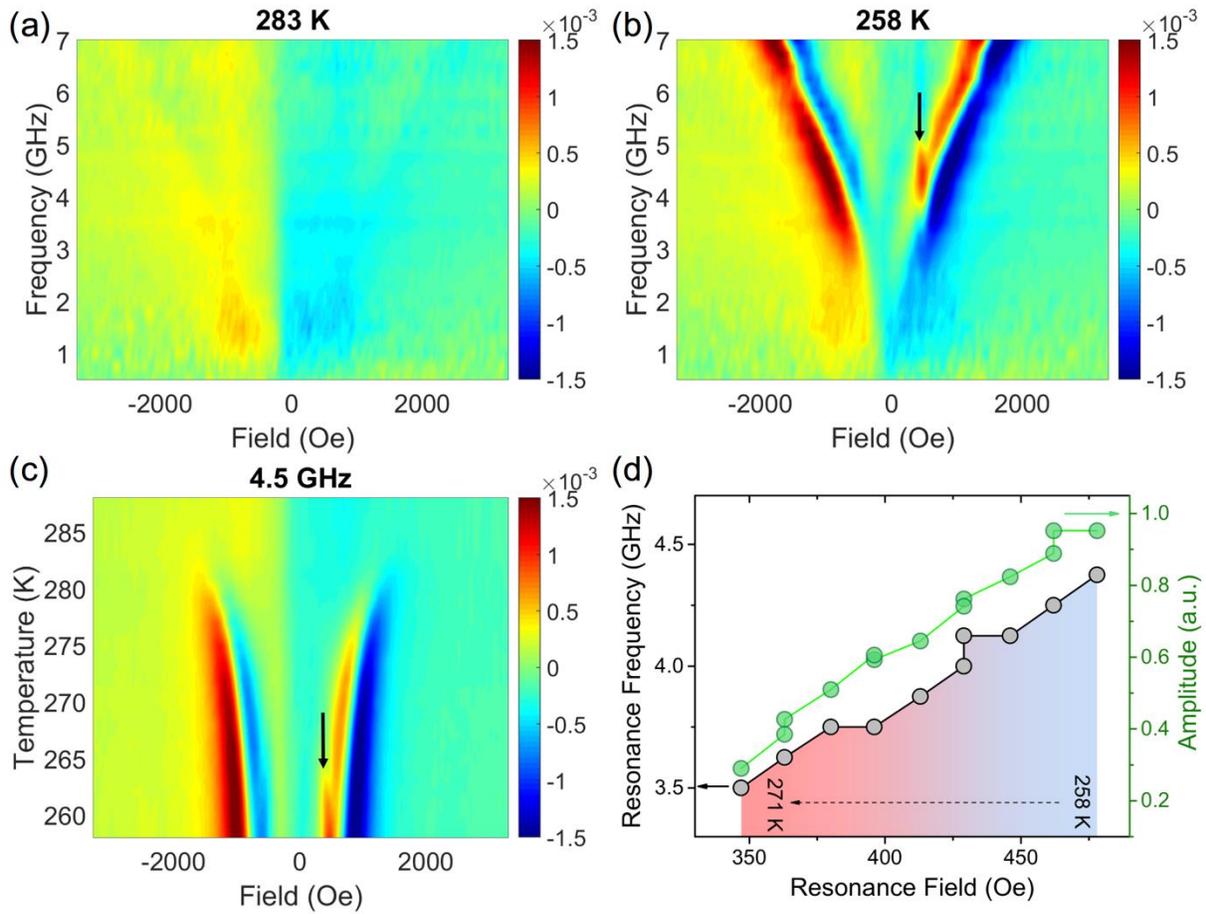

**FIG 4** Experimental measurements of microwave absorption spectroscopy (MAS) of FeGe film at different temperature, in-plane magnetic field, and microwave frequency. a) shows the MAS above the critical temperature with no clear absorption feature. b) shows the MAS below the critical temperature, which has the helical (arrow) and field-polarized state resonances. c) the MAS at 4.5 GHz as the field and the temperature vary. (d) shows trend of the frequency, field, and amplitude of the helical resonance for each temperature.

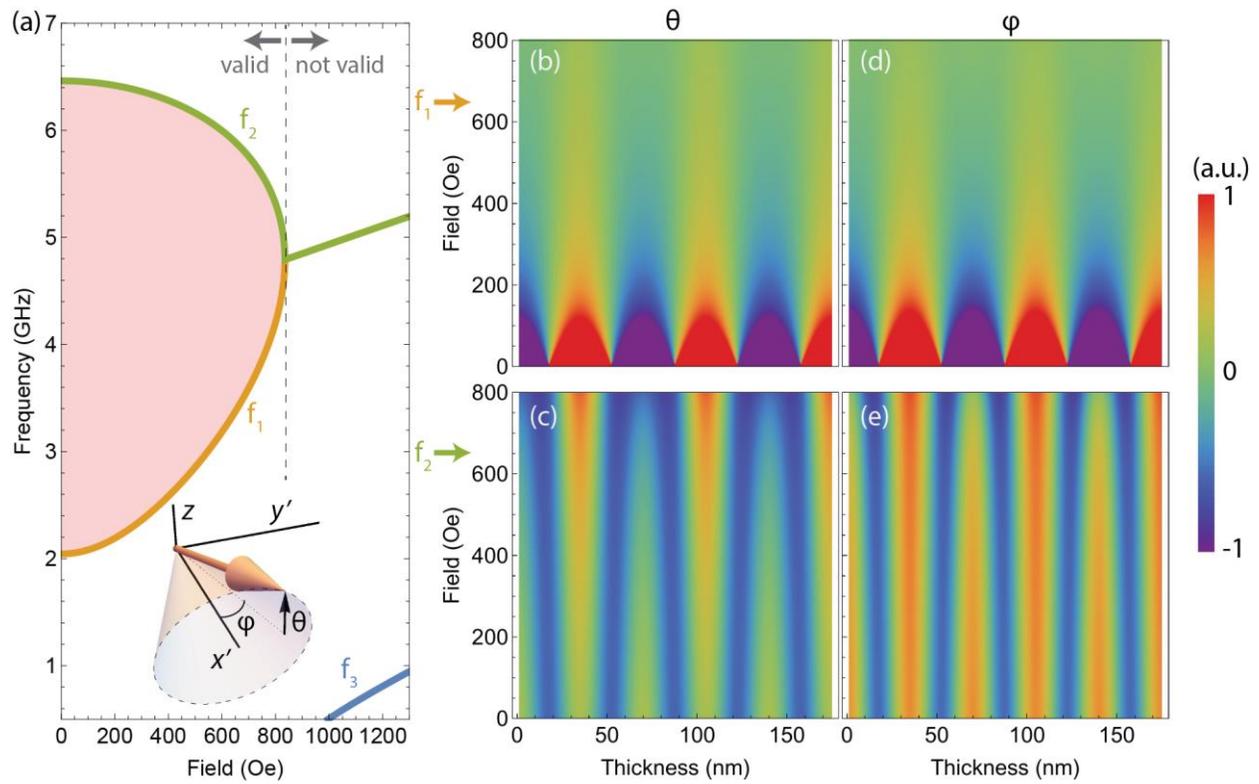

**FIG 5** Analytical calculation of the resonance frequencies and oscillation modes. a) the real parts of the eigenvalues of the equations of motion result in three resonance frequencies $f_1$, $f_2$, and $f_3$. The inset shows the coordinate system and angle variables of spins. Because of non-zero imaginary parts of the eigenvalues, only the pink-filled region represents correct resonance feature. b) and c) show the oscillation modes of $\theta$ (out-of-plane) for $f_1$ and $f_2$, respectively. d) and e) show the oscillation modes of $\varphi$ (in-plane) for $f_1$ and $f_2$ frequencies, respectively.

# Supplemental Material: Chiral magnetic excitations in FeGe films

Emrah Turgut[1], Albert Park[1], Kayla Nguyen[1], Austin Moehle[1], David A. Muller[1,2], and Gregory D. Fuchs[1,2]

1. School of Applied and Engineering Physics, Cornell University, Ithaca, NY 14853, USA
2. Kavli Institute at Cornell for Nanoscale Science, Ithaca, NY 14853, USA

This Supplemental Material (SM) provides additional information about the sample characterization, microwave absorption spectroscopy measurements, computational and analytic calculations.

### S1. Cross section scanning electron micrograph of the FeGe film

To simulate micromagnetics accurately, we precisely measure the thickness of our film. This is essential because standing spin wave modes and frequencies strongly depend on the film thickness. We used a focused ion beam to make a cross section cut of the film and scanning electron microscopy to measure the thickness. In Fig. S1, we show the cross section images of the film and measurement of the thickness as 176 nm.

### S2. X-ray diffraction and lattice parameters

From the X-ray diffraction measurements in Fig. 1a, we found $d_{111}$ 2.702(1) Å for FeGe and 3.141(1) Å for Si from the corresponding peaks. The vertical lattice constants are then found to be 4.680 Å for FeGe and 5.441 Å for the Si substrate. The reported lattice constant for the bulk FeGe is 4.701 Å at 290 K [1]. The lattice mismatch becomes

$$\frac{4.701 - 5.441 \cos 30}{4.701} = -0.23 \, \%.$$

This lattice mismatch would be released at every (23/10000≈1/435) 435 lattice sites that equals to 204 nm. Indeed, our electron backscattering diffraction micrograph indicates an average 300 nm grain size.

To find the stress on the FeGe film, we use Murnaghan formula [2], which is

$$\frac{V_0}{V} = (1 + kP)^{1/ck}$$

where $V_0$ is the equilibrium and $V$ is the final volume, $P$ is the pressure, c is the bulk modulus, i.e. $c = -V\frac{dp}{dV}$, ck is the derivative, i.e. $ck = -\frac{d}{dp}\left(V\frac{dp}{dV}\right)$. Solving using these formulas numerically, we find the pressure -62 MPa.

To understand crystallization process in FeGe by post-growth annealing, we calculate thermal expansion in FeGe and Si lattices. Thermal expansion coefficients in FeGe and Si were reported $1.7 \times 10^{-5}$ and $3.6 \times 10^{-6}$ $C^{-1}$ respectively [3,4]. At temperature 192 °C, the lattice constants theoretically match, and at higher temperatures Fe and Ge atoms form into the B20 crystal configuration. In our FeGe films, we use relatively high temperature (350 °C) to form B20. Indeed, lower temperature annealing produces weaker FeGe [111] peak in the x-ray diffraction. The recent study on FeGe thin films also reported 290 °C is the optimum annealing temperature [5].

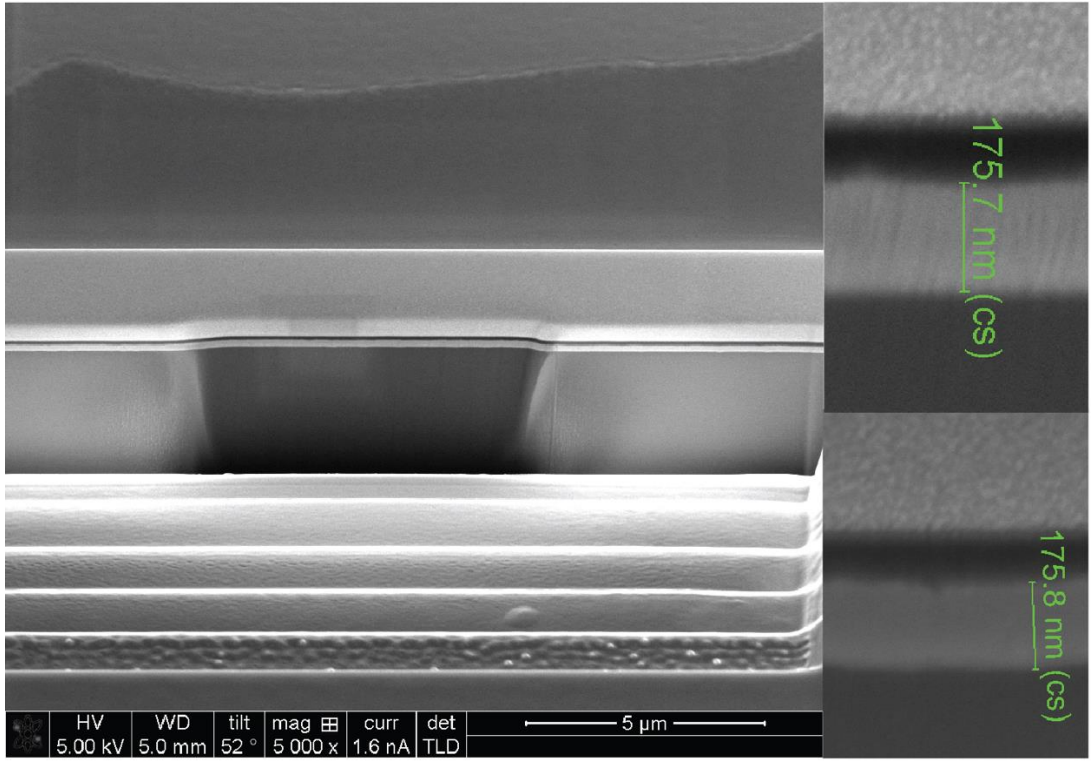

**FIG S1. SEM cross section images of the film. The focused ion beam is used to make a cross section and the scanning electron microscopy reveals the thickness of 176 nm.**

**S3. AC Magnetic susceptibility**

In this section, we show the measurements of the critical temperature of FeGe film determined with AC magnetic susceptibility using a Quantum Design Physical Property Measurement System (Fig. S2).

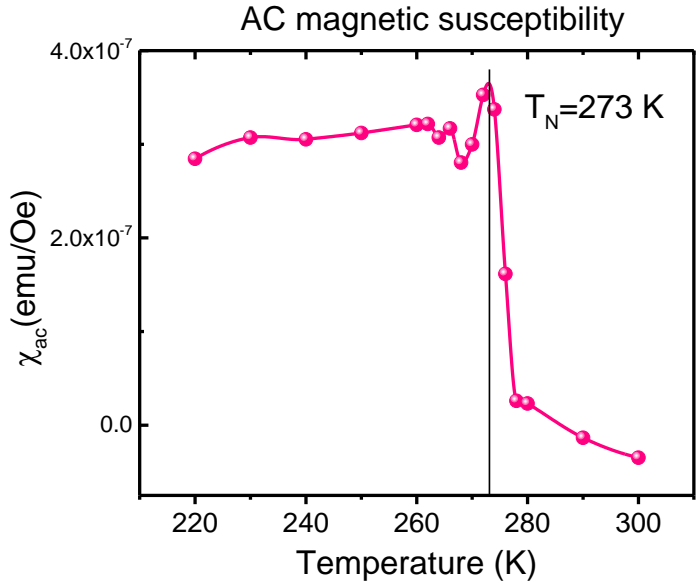

**FIG S2.** AC magnetic susceptibility measurements of FeGe film with a 20 Oe constant field and a 10 Oe ac excitation field at 25 Hz. The critical temperature is obtained 273 K, which is slightly lower than the bulk value of 278 K.

### S4. Parameters used in the micromagnetic simulations

Here we show an example input file for the Mumax3 micromagnetic simulation. One can find used material parameters in the "Material parameter" section.

```
Nx :=4
Ny :=4
Nz :=220  // number of cells in x, y, and z directions
c := 8e-10 // cell size in nm
SetMesh(Nx, Ny, Nz, 8e-10, 8e-10, 8e-10, 16, 16, 0) // Setting
mesh with 16 periodic boundary conditions in x and y
lmd := 70e-9  // helix period in nm
mask := newVectorMask(Nx, Ny, Nz)

for i:=0; i<Nx; i++{
    for j:=0; j<Ny; j++{
        for k:=0; k<Nz; k++{
        xdir := cos(2*pi*k*c/lmd)
        ydir := sin(2*pi*k*c/lmd)
        zdir := 0
```

```
            mask.setVector(i, j, k, vector(xdir, ydir, zdir))
            }
        }
    }
}
m.SetArray(mask)
save(m)
// Material parameters
Ku1 = -3500 // uniaxial anisotropy
Msat = 150e3 // Saturation magnetization
exccons := 6.8e-13 // Exchange constant
Aex = exccons
DD := 4*pi*exccons/lmd // Bulk DMI constant
Dbulk = DD
Kc1 = 0 // crystalline anisotrpy
anisc1 = vector(1,1,1)
anisU = vector(0, 0, 1) // vestors for anisotropies
B := -0.05 // Magnitude of the external magnetic field in Tesla
alpha = 0.002 //the damping constant
B_ext = vector(B, 0, 0) // External magnetic field vector
relax() // Finding the equilimbrium
save(m)
TableAdd(B_ext)
tableAdd(E_total)
tableautosave(5e-13)
autosave(m, 2.5e-11) // saving magnetization at every 25 ps.
t0 :=2e-9 // 'ringdown' magnetic excitation time zero
fwhm := 1e-10 // full width at half maximum
A := 0.0005 // amplitude of the excitation in Tesla

B_ext  =  vector(B,  A*exp(-(t-t0)*(t-t0)/pow(fwhm,2)),  0)  //
external field during excitation
run(20e-9) // 20 ns run time.
save(m)
```

**S5. Additional microwave absorption spectrum**

In the main text, we show the microwave absorption spectrum only at 283 and 258 K, and at 4.5 GHz frequency. Here in Fig. S3 and S4, we show the rest of spectrum at other temperatures and microwave frequencies.

**S6. Kittle's fit and linewidth analysis of microwave absorption spectrum**

We describe our microwave absorption spectrum lineshaps with a Lorentzian shape. Because we measure the derivative of the spectrum, we fit the lineshapes with

$$f(H) = a + b \frac{(H-H_0)}{[(\Delta H)^2 + (H-H_0)^2]^2} + c\,H, \tag{S3.1}$$

where *a*, *b*, and *c* are arbitrary coefficients, $H_0$ is the resonance field, and $\Delta H$ is the linewidth. In Fig. S5, we show dependence of the linewidths on frequency at 258 and 263 K. Because the field-polarized states can be described by Kittel's formula, ΔH depends on Gilbert damping α by

$$\Delta H = \Delta H_0 + \frac{4\pi \alpha f}{\gamma_e}, \tag{S3.2}$$

where $\Delta H_0$ is frequency independent broadening component, and $\gamma_e$ is the electron's gyromagnetic ratio. After we fit linewidths with Eq. S2, we find Gilbert damping $\alpha = 0.038 \pm 0.005$. This is significantly lower than recently reported 0.28 damping constant with an out-of-plane magnetic field along [111] direction [6].

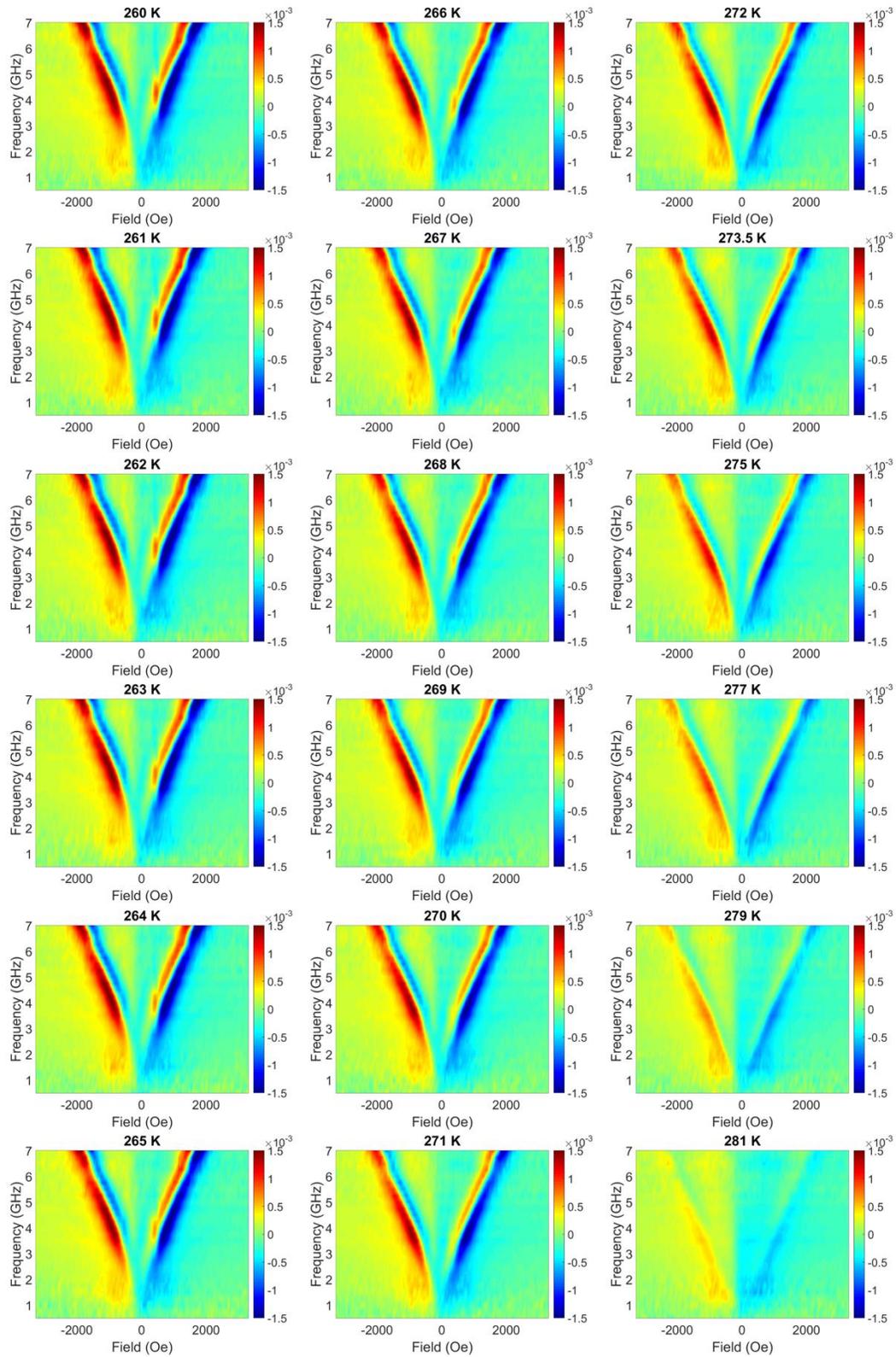

**Fig S3. Microwave absorption spectrum of FeGe film at different temperatures.**

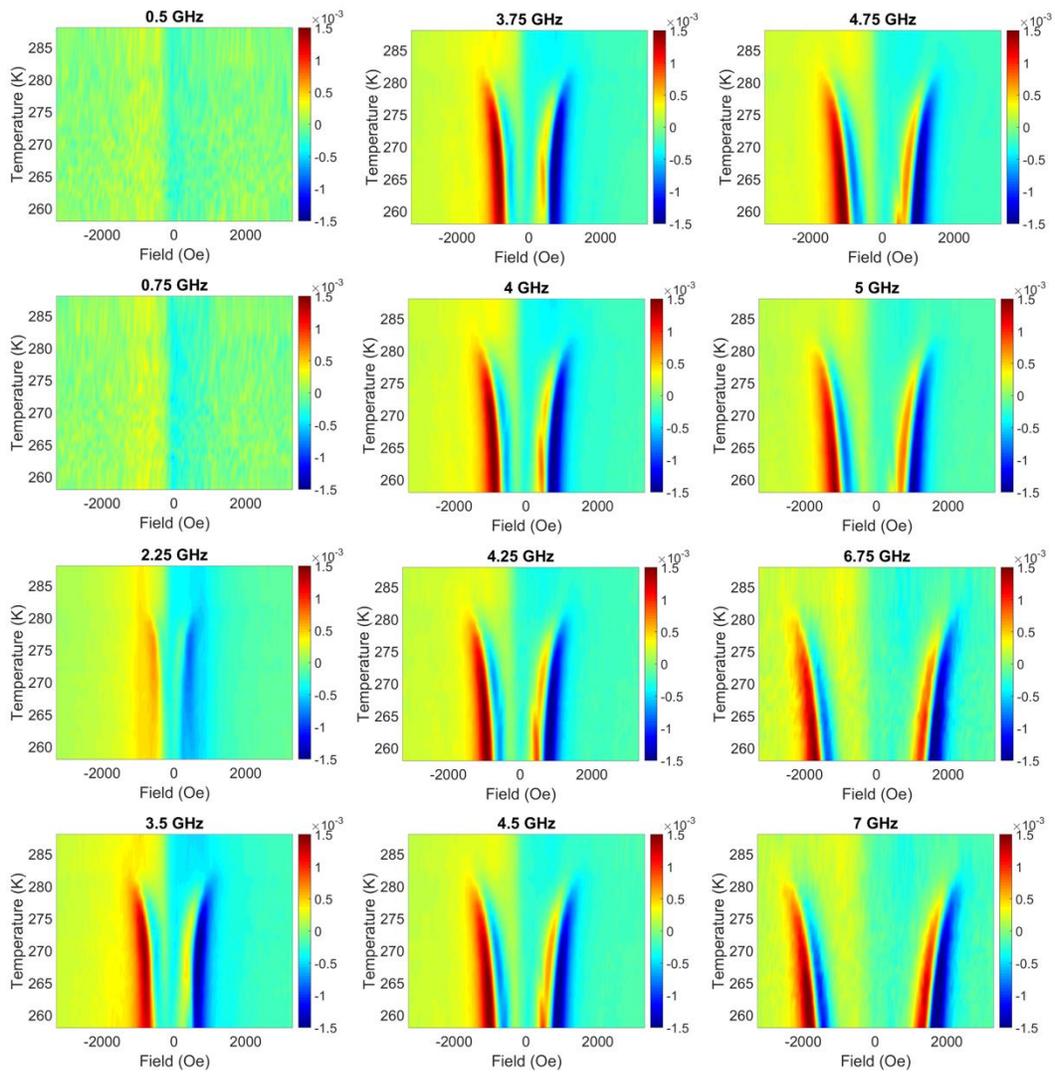

**FIG S4. Microwave absorption spectrum of FeGe film at different microwave frequencies.**

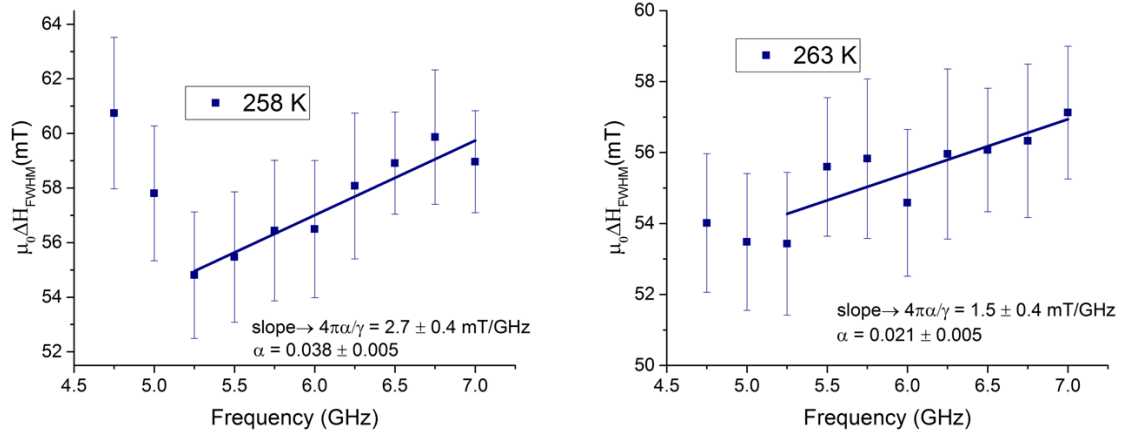

**FIG S5. Dependence of the linewidth to the microwave frequency at 258 K and 263 K temperatures. We find Gilbert's damping constant 0.038 and 0.021 at these temperatures. The first couple of data points have a contribution from helical resonance, we take data points between 5.25 GHz and 7 GHz into account.**

**S7. Details of analytic calculation**

We describe the one dimensional Hamiltonian density of a chiral helimagnet by

$$\mathcal{H} = A(\partial_z \boldsymbol{m})^2 - \boldsymbol{D}.\boldsymbol{m} \times \partial_z \boldsymbol{m} - \boldsymbol{B}.\boldsymbol{m} + K_u(\boldsymbol{m}.\widehat{\boldsymbol{n}})^2 - \frac{1}{2}\boldsymbol{H_m}.\boldsymbol{m}, \quad (S5.1)$$

where $A$ is the exchange stiffness constant, $\boldsymbol{D}$ is the DM interaction constant, $\boldsymbol{B}$ is the external magnetic field, $K_u$ is the anisotropy constant and $\boldsymbol{H_m}$ is the demagnetizing field due to shape of the sample. We define the normalized magnetization $m = [\sin\theta(z,t)\cos\phi(z,t), \sin\theta(z,t)\sin\phi(z,t), \cos\theta(z,t)]$ in spherical coordinates, where $\theta(z,t)$ and $\phi(z,t)$ are the polar angles and $z$ is the normal axis of the film. The external magnetic field $\boldsymbol{B}$ includes dc field $B_x$ and ac microwave field $B_y$, and is written as $\boldsymbol{B} = [B_x, B_y \sin\omega t, 0]$. Next, we write down the Lagrangian density $\mathcal{L}$,

$$\mathcal{L} = -\frac{\hbar M_s}{g_e \mu_B}(\cos\theta - 1)\partial_t\phi - \mathcal{H}, \quad (S5.2)$$

and the Rayleigh dissipation $\mathfrak{R} = \frac{\alpha}{2}(\partial_t \boldsymbol{m})^2$, where $\alpha$ is the Gilbert damping parameter, $M_s$ is the magnetization, $\mu_B$ is the Bohr magnetron, $g_e$ is the electron g-factor. Next, the equations of motion are constructed from Eq. S5.2 in terms of $\theta(z,t)$ and $\phi(z,t)$ using

$$\frac{d}{dt}\frac{\delta \mathcal{L}}{\delta(\partial_t \theta)} - \frac{\delta \mathcal{L}}{\delta \theta} + \frac{\delta \mathcal{R}}{\delta(\partial_t \theta)} = 0, \tag{S5.3}$$

$$\frac{d}{dt}\frac{\delta \mathcal{L}}{\delta(\partial_t \phi)} - \frac{\delta \mathcal{L}}{\delta \phi} + \frac{\delta \mathcal{R}}{\delta(\partial_t \phi)} = 0, \tag{S5.4}$$

$$L_m \sin\theta\, \partial_t\phi = \sin 2\theta\, \partial_z\phi(-D + A\partial_z\phi) - \cos\theta\left(B_x \cos\phi + B_y \sin\phi \sin\omega t\right) - 2A\partial_z^2\theta, \tag{S5.5}$$

$$2 L_m \partial_t\theta = B_y \sin\omega t \cos\phi - B_x \sin\phi + 2\cos\theta\, \partial_z\theta(-D + 2A\partial_z\phi) + 2A\sin\theta\, \partial_z^2\phi. \tag{S5.6}$$

where $L_m = \frac{\hbar M_s}{g_e \mu_B}$.

Then, we substitute following approximation for $\phi_1$ and $\theta_1$ by using small angle approximation:

$$\phi(z,t) = Qz + \phi_1(z) \sin\omega t, \tag{S5.7}$$

$$\theta(z,t) = \frac{\pi}{2} + \theta_1(z) \cos\omega t, \tag{S5.8}$$

$$\phi_1(z) = A_1 + A_2 \cos Qz + A_3 \cos^2 Qz \tag{S5.9}$$

$$\theta_1(z) = B_1 + B_2 \cos Qz + B_3 \cos^2 Qz \tag{S5.10}$$

Here, we use two methods: (1) With $B_y=0$, we find the natural frequencies by solving the eigenvalue problem, and (2) we use a non-zero $B_y$ and solve 'driven harmonic oscillator' problem.

For the first method

$$\boldsymbol{I}\,\omega^2 - \boldsymbol{\mathcal{M}} = 0, \tag{S5.11}$$

where $\boldsymbol{I}$ is the identity matrix and M is

$$\frac{1}{2L_m^2}\begin{pmatrix} 0 & 4 B_x A Q^2 & 40 A^2 Q^4 \\ -4 A B_x Q^2 & -8 A^2 Q^4 & 16 A B_x Q^2 \\ -B_x^2 & -12 A B_x Q^2 & -80 A^2 Q^4 \end{pmatrix}. \tag{S5.12}$$

For the second method, A and B coefficients have common denominator

$$40A^4 B_x^2 Q^8 - 8A^2 B_x^4 Q^4 + (62 B_x^2 Q^4 + 160 A^2 Q^8) L_m^2 A^2 \omega^2 - 44 A^2 Q^4 L_m^4 \omega^4 + L_m^6 \omega^6. \tag{S5.13}$$

Eq. S5.13 must be zero for the resonance phenomena to give large oscillation amplitude. This also gives the same frequencies as in Fig. 5 in the main text.

Moreover, we show the imaginary parts of the resonance frequencies (Fig. S5) to clarify why the pink-filled region in the Fig. 5 is the correct description for resonance. In Fig. S5, we show the imaginary parts of the eigenvalues at $f_1$ and $f_2$, which have zero imaginary parts below 830 Oe magnetic field and non-zero above it. Thus, the solutions for higher fields are not valid.

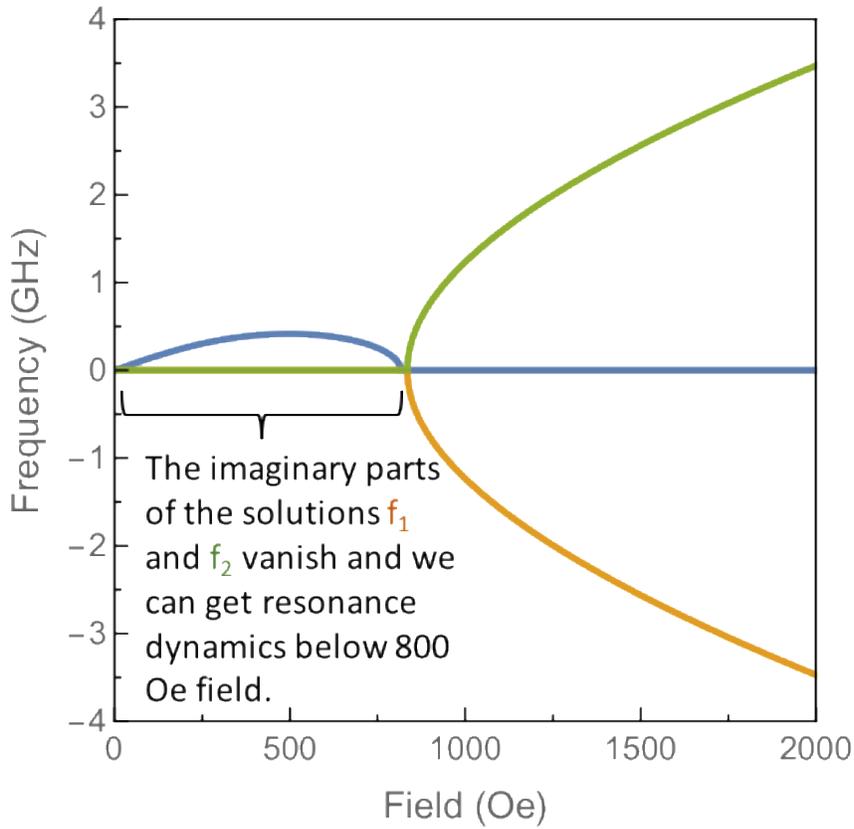

**FIG S6. The imaginary parts of the eigenvalues of $f_1$ and $f_2$ frequency solutions. Because of non-zero imaginary parts above 830 Oe, only the low field solutions are valid. This also confirms the small angle approximation made in the analytic model.**